\documentclass[runningheads]{llncs}
\usepackage{graphicx}
\usepackage{amsmath,amssymb} 
\usepackage{color}
\usepackage[width=135mm,left=23mm,paperwidth=176mm,height=210mm,top=18mm,paperheight=246mm]{geometry}
\usepackage{orcidlink}

\begin{document}

\setlength{\textfloatsep}{6pt plus 2pt minus 2pt}

\title{A Lean and Spec-Driven AI-Assisted Software Development Lifecycle
for Applied AI Education: \texorpdfstring{\\}{ }The AI-SDLC Approach}

\titlerunning{The AI-SDLC Approach}
\authorrunning{A. Martin and S. Schwander}

\author{Andreas Martin \orcidlink{0000-0002-7909-7663} \and
        Sandro Schwander \orcidlink{0009-0006-9708-3800}}

\institute{FHNW University of Applied Sciences and Arts Northwestern Switzerland,\\
School of Business, Riggenbachstrasse 16, 4600 Olten, Switzerland\\
\email{andreas.martin@fhnw.ch}, \email{sandro.schwander@fhnw.ch}}

\maketitle

\begin{abstract}
AI coding agents increasingly support software development beyond code completion, including planning, implementation, testing, and repository-level task execution. Their practical use, however, often remains only weakly connected to established software engineering practices. The aim of this work is to develop and evaluate a lightweight, spec-driven lifecycle for governed agentic software engineering. The lifecycle combines established software engineering practices with repository-local guidance through specifications, AGENTS.md, and phase-specific agent skill files. The approach was developed in the context of the FHNW course AI-assisted Software Development and applied by students to business-oriented software use cases. Its educational and practical applicability is explored through a student survey combining closed rating items with open-ended questions. The contribution of this work is a process-oriented framework that enables AI coding agents to operate with bounded autonomy within an explicit, reviewable, and test-oriented software development lifecycle.
\end{abstract}

\section{Introduction}

AI coding agents have become part of everyday software development. Generating code is no longer their most distinctive capability. Contemporary systems can interpret requirements, propose implementation strategies, generate tests, navigate repositories, and modify existing software artifacts \cite{sapkota_vibe_2025}, \cite{dong_survey_2025}, \cite{madani_towards_2025}. As a result, the role of the developer is beginning to change.

The practical implications of this shift remain only partially understood. Much of the current literature concentrates on implementation performance, benchmark results, and coding productivity \cite{madani_towards_2025}, \cite{wang_software_2025}. Software development, however, involves more than implementation. Requirements must remain traceable, architectural decisions must remain coherent over time, and software artifacts must pass validation and deployment processes before they can be considered operational. Several authors have observed that current research and evaluation practices place considerably greater emphasis on implementation and testing than on requirements engineering, software design, deployment, or lifecycle governance \cite{wang_software_2025}. Empirical studies of agentic coding point in a similar direction. AI-generated contributions can support development activities effectively, yet human review remains essential for quality assurance and contextual adaptation \cite{watanabe_use_2026}, \cite{abrahao_software_2025}.

Similar observations emerged during the development of the FHNW course AI-assisted Software Development. Students were generally able to produce functional software with modern coding agents after a relatively short familiarization period. Difficulties appeared elsewhere. As projects evolved, requirements, architectural decisions, tests, and implementation artifacts increasingly drifted apart. Development activities often converged toward prompt-driven implementation, while the underlying rationale behind development decisions remained implicit and difficult to reconstruct.

The AI-Assisted Software Development Life Cycle (AI-SDLC) originated from this observation. The initial objective was not to develop another software engineering methodology. A more practical concern motivated the work. Students needed a lightweight process structure that would allow them to use AI coding agents productively without abandoning the engineering practices required for maintainable, testable, and deployable software systems.

The relevance of this problem extends beyond educational settings. Security-oriented studies suggest that apparently plausible and functionally correct agent-generated code may still contain vulnerabilities and quality deficiencies that are difficult to detect without systematic validation procedures \cite{zhao_is_2025}. At the same time, software teams increasingly experiment with agentic workflows that extend beyond isolated coding tasks. Under these conditions, productivity gains alone provide an insufficient basis for evaluation. The relationship between specifications, architecture, testing, deployment, and human oversight becomes equally important.

The purpose of this study is to develop and explore the applicability of the AI-SDLC as a lightweight lifecycle for more disciplined AI-assisted software development. The approach was applied in the FHNW course AI-assisted Software Development, where student teams used the lifecycle while developing business-oriented software solutions with contemporary coding agents. Following project completion, participants completed a short cross-sectional survey addressing process support, usability of the lifecycle artifacts, perceived usefulness of repository-local self-instruction, and the interaction between AI-generated outputs and human review activities.

\section{AI-Assisted Software Development Lifecycle (AI-SDLC)}

AI-SDLC represents one response to the challenges outlined above. The lightweight approach combines established software engineering practices, including Test-Driven Development (TDD) \cite{beck_test-driven_2003}, the testing pyramid \cite{fowler_test_2012}, Clean Architecture \cite{martin_clean_2017}, CI/CD with continuous integration \cite{fowler_continuous_2006}, containerized validation and continuous deployment with DevOps \cite{humble_continuous_2010,forsgren_accelerate_2018}, Kanban-based execution \cite{anderson_kanban_2010}, and selected concepts from the Scaled Agile Framework (SAFe) \cite{leffingwell_scaling_2007,knaster2020safe}, within a lightweight lifecycle structure, together with more recent approaches to spec-driven development \cite{piskala_spec-driven_2026,github_specification-driven_2025,bockeler_understanding_2025}.

\begin{figure}[t]
\centering
\includegraphics[width=\linewidth]{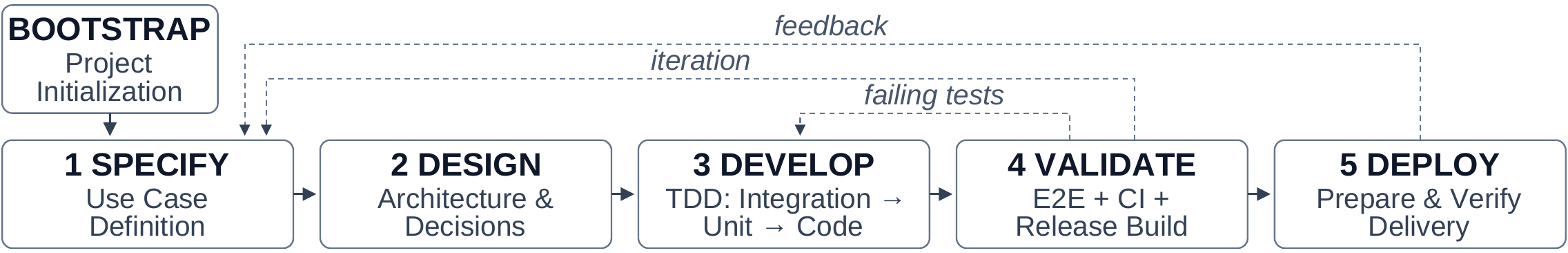}
\caption{Phases of the AI-Assisted Software Development Life Cycle (AI-SDLC)}
\label{fig:AI-SDLC}
\end{figure}

As depicted in Fig.~\ref{fig:AI-SDLC}, the activities are organized into six phases: Bootstrap, Specify, Design, Develop, Validate, and Deploy. These phases are intentionally minimal. Their purpose is to provide coordination, traceability, and lifecycle awareness rather than procedural rigidity. The AI-SDLC artifacts described in this section are available as a versioned software artifact on Zenodo \cite{martin_ai-sdlc_2026-1}.

Instead of relying exclusively on conversational prompts, AI-SDLC embeds development instructions directly into project artifacts. The \texttt{AGENTS.md} file \cite{agentsmd_agentsmd_2026} provides lifecycle guidance and operational constraints for coding agents. Additional \texttt{SKILL.md} files \cite{agent_skills_agent_2026} define phase-specific behavior, while specifications, task-state documents, and architecture artifacts provide contextual grounding. Together, these elements establish a form of repository-local self-instruction that remains visible, reviewable, and reusable throughout the project lifecycle.

\subsection{Lifecycle Phases}

The lifecycle starts with \textit{Bootstrap}, which establishes the project context, development environment, and architectural baseline. Relevant information about the system, technology stack, architecture, and dependencies is maintained in \texttt{docs/PROJECT.md}. \textit{Specify} then transforms a user story or change request into an explicit use case specification under \texttt{docs/specs/}, including intended behavior, scope, acceptance criteria, non-functional requirements, and intended tests. No implementation is performed during this phase.

During \textit{Design}, the specification is mapped to the architecture following Clean Architecture. Required domain and application behavior, interfaces, infrastructure components, and dependencies are identified, consequential decisions can be documented as Architecture Decision Records (ADRs), and the implementation work is divided into tasks in \texttt{docs/TASKS.md}. \textit{Develop} implements these tasks according to TDD, starting with integration and unit tests before producing the corresponding implementation through the Red--Green--Refactor cycle.

\textit{Validate} considers the complete use case through automated unit, integration, and, where appropriate, end-to-end tests, together with CI and validation of the release artifact. Failed validation can return development to an earlier phase. Finally, \textit{Deploy} prepares and verifies delivery of the validated artifact and, where authorized and configured, executes deployment to the target environment. Operational feedback or changed requirements can initiate another lifecycle iteration. AI-SDLC is therefore iterative rather than a linear or waterfall process.

\subsection{Repository-Local Guidance for AI Coding Agents}

A central characteristic of AI-SDLC is that development guidance and project context are maintained as version-controlled repository artifacts. Repository-level instruction files such as \texttt{AGENTS.md} provide an open format for supplying coding agents with persistent project-specific instructions \cite{agentsmd_agentsmd_2026}. Gloaguen et al. \cite{gloaguen_evaluating_2026} found that coding agents generally follow such instructions, while larger context files do not necessarily improve task success and can increase inference costs.

Accordingly, AI-SDLC keeps \texttt{AGENTS.md} compact and uses it as the central entry point containing common workflow rules and guardrails. The current lifecycle state and tasks are maintained in \texttt{docs/TASKS.md}, architectural and runtime context in \texttt{docs/PROJECT.md}, and intended behavior in use case specifications under \texttt{docs/specs/}. Detailed procedural instructions are separated into phase-specific Agent Skills \cite{agent_skills_agent_2026}, each represented by a \texttt{SKILL.md} file under \texttt{skills/ai-sdlc-*}.

Consequential architectural decisions can additionally be recorded as Architecture Decision Records (ADRs), preserving their context, decision, and consequences \cite{nygard_documenting_2011}. Together, these artifacts separate general process guidance, phase-specific procedures, project context, lifecycle state, specifications, and architectural decisions. They persist across agent sessions and can be inspected, modified, and versioned together with source code and tests.

\subsection{Human Control and Agent Autonomy}

AI-SDLC allows coding agents to operate with substantial autonomy within individual lifecycle phases. Once a phase has been initiated and the required input artifacts are available, the agent can perform the activities defined by the corresponding skill, use available development tools, and create or modify the required artifacts without requiring human approval after every individual action.

Human control is maintained through the persistent artifacts connecting the lifecycle phases. Specifications, architectural decisions, tasks, tests, implementation artifacts, and validation results remain inspectable and can be modified whenever they do not adequately represent the intended requirements, architecture, or quality expectations. Modified artifacts subsequently become part of the agent's context for further activities. AI-SDLC thereby combines autonomous execution within lifecycle phases with persistent and reviewable artifacts through which developers retain control over the evolving software system.

\section{Evaluation Method}

We conducted a cross-sectional online survey among the students of a single course cohort. The questionnaire combined closed rating items with open-ended questions, so that the ratings could be read together with the situations respondents described in their own words. Since the study covers one cohort at one point in time and has no control group, it is descriptive and exploratory and is not designed to establish causal effects.

Participants were students of the bachelor's program Business Artificial Intelligence who had taken the module ``AI-Assisted Software Development'' and had carried out a
semester project with the AI-SDLC. The version of AI-SDLC evaluated in the module is archived separately on Zenodo \cite{martin_ai-sdlc_2026}. All students of the module were invited to take part. The invitation was issued after the module had been completed and graded, so that participation could not affect assessment. Participation was voluntary, and the survey link was not personalised, so responses cannot be traced back to individual students. The questionnaire was implemented in German. Of the 34 invited students, 23 opened the
questionnaire and 13 submitted it completely, which corresponds to a response rate of 38\% and, among those who started, a completion rate of 57\%. Completeness was
determined from the platform's disposition codes. The median time to complete the questionnaire was about 22 minutes.

The respondents' median age was 25 years (range 23--32). Programming experience was limited: five respondents described themselves as beginners and seven as having basic knowledge, with a median of three years of programming including studies and hobby projects, and prior development work had taken place mainly in hobby or open-source projects (10 of 13). Experience with AI coding tools, in contrast, was high: seven respondents had used such tools daily or intensively before
the module, and none reported no experience at all.

\subsection{Survey Structure}

The questionnaire comprised six sections (A--F). Sections A and B collected participant background, prior software development experience, use of AI coding tools, and familiarity with software engineering concepts. Section C addressed the AI-SDLC in five thematic blocks covering process support, lifecycle artifacts, repository-local instructions, human verification, and learning and productivity. Statements were rated on a five-point agreement scale with an additional ``not applicable / don't know'' option. Each block contained at least one reverse-worded item to counteract acquiescence bias \cite{weijters2012misresponse}, and blocks C-I to C-IV included open-ended questions asking respondents to describe concrete situations from their projects.

Section D separately measured adherence to and perceived usefulness of each lifecycle phase, following the distinction between adherence and perceived value \cite{carroll2007fidelity,davis1989perceived}. Section E contained four open-ended questions on benefits, review practices, improvements, and further remarks. Section F asked respondents how likely they would be to recommend AI-SDLC to a fellow student on a scale from 0 to 10 \cite{reichheld2003one}.

\subsection{Analysis}

All closed items were analyzed descriptively. Given the ordinal scales and the small sample, we report frequency distributions and absolute counts and use no inferential statistics. Reverse-worded items were retained in their original direction, and no composite scores were formed. Open-ended responses were coded thematically by one author; categories may overlap.

\begin{table}[!t]
\centering
\small
\renewcommand{\arraystretch}{0.9}
\caption{Structure of the survey instrument}
\label{table:parts_questionnaire}
\begin{tabular}{ll}
\hline
Part & Content \\
\hline
A) General questions & Age, degree program, current semester, experience \\
B) Previous experience & Contexts, process models, AI tools, familiarity with terms \\
C-I) Process support & Statements on the AI-SDLC as a whole \\
C-II) Lifecycle artifacts & Statements on user stories, architecture, specifications \\
C-III) Self-instruction files & Statements on AGENTS.md, skills, project-local rules \\
C-IV) Human verification & Statements on review, TDD and traceability \\
C-V) Learning and productivity & Statements on learning, speed and reuse \\
D) Phases & Adherence and perceived usefulness per phase \\
E) General feedback & Benefits, review practice, suggestions, remarks \\
F) Recommendation & Net Promoter Score item, consent for follow-up \\
\hline
\end{tabular}
\end{table}

\section{Findings}

\begin{figure}[t]
\centering
\includegraphics[width=\linewidth]{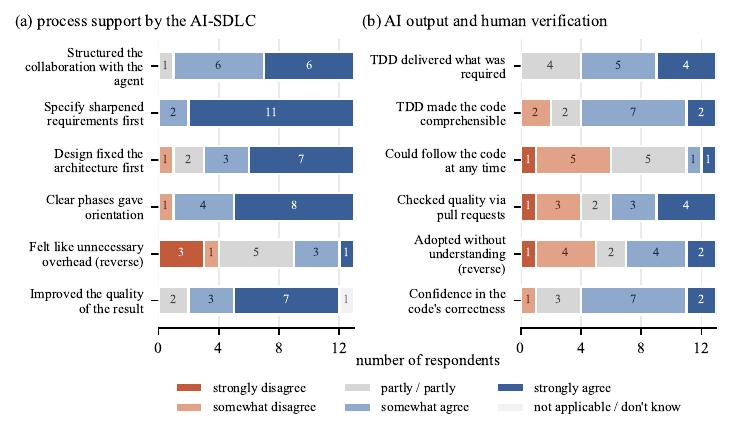}
\caption{Agreement with (a) the statements on process support by the AI-SDLC
         and (b) AI output and human verification ($n=13$).}
\label{fig:likert}
\end{figure}

Overall, the AI-SDLC received broad approval (Fig.~\ref{fig:likert}a). The clearest result concerns the Specify step: all 13 respondents agreed that it led them to sharpen the requirements before the AI started implementing, 11 of them strongly. Twelve of 13 agreed that the process helped structure the collaboration with the AI agent, and the same number that the clear sequence of phases gave them orientation in the project. Ten of 13 stated that the structured process improved the quality of their project result compared to their usual way of working. The open answers name goal-oriented feature design, clear guidelines and the step-by-step approach as the reasons; one respondent reported that the strict separation of Design and Validate prevented the agent from producing faulty code during a final refactoring. The counterweight is effort: four respondents agreed that the AI-SDLC felt like unnecessary overhead in individual phases, the free-text answers locate this criticism specifically in small tasks, where several respondents found the process disproportionate to the change being made.

Repository-local instruction files were judged positively throughout. Eleven of 13 respondents found it clear how to use and adapt such files, nine of them strongly. The reverse-worded control item points the same way: not a single respondent agreed that the files had hardly any noticeable influence, and 11 of 13 rejected the statement outright. AGENTS.md was the most widely used file (10 of 13), followed by phase-specific skill files (5 of 13).

Human verification of AI output shows a more ambivalent picture (Fig.~\ref{fig:likert}b). Respondents largely trusted the result: test-driven development kept the agent on what was actually required, and most were confident the code was correct, though mostly with reservations. This trust rested on the tests rather than on understanding the code.
Following the agent's code at any time was one of the lowest-rated items in the survey, and only about half checked quality actively through pull requests. The free-text answers make this gap explicit: while six respondents admitted on the closed item to adopting suggestions without really understanding them, 11 of 13 described such a situation in their own words, citing fatigue in long sessions, time pressure, and features beyond their own programming skills. Verification thus shifted from reviewing code to trusting tests.

\begin{figure}[t]
\centering
\includegraphics[width=\linewidth]{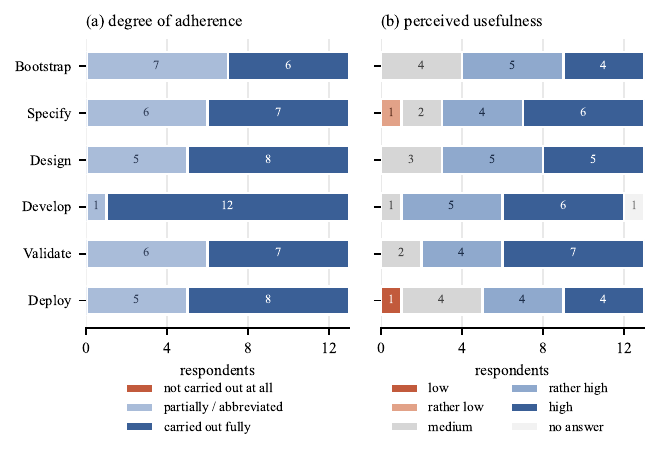}
\caption{Degree of adherence (a) and perceived usefulness (b) per lifecycle
phase ($n = 13$), both in process order. No respondent reported skipping a
phase entirely, which is why the level ``not at all'' appears in the legend
of panel (a) without a corresponding bar.}
\label{fig:phases}
\end{figure}

The same pressures shaped how the process itself was followed (Fig.~\ref{fig:phases}). No phase was skipped, but only Develop was carried out fully by almost everyone, while
Bootstrap, Specify and Validate were shortened most often (Fig.~\ref{fig:phases}a). This was not a verdict on their value: every phase was rated useful, with means between 3.77 and 4.42 on the five-point scale (Fig.~\ref{fig:phases}b). Eight of the nine respondents asked about it described shortening a phase, citing time pressure, effort, deliberate scope decisions for low-risk tasks, impatience at the start of the project, and in one case a lack of understanding. Bootstrap is the clearest case: as a one-time setup at the start, its effort comes before any visible benefit.

Asked about learning and productivity, all 13 respondents agreed that they made faster progress with the AI-SDLC, nine of them strongly, and 12 of 13 would use the process again voluntarily in a future project. Eleven of 12 agreed that it helped them understand software engineering concepts better. The one item that divides the group is self-directed learning: five respondents agreed that they had learned less themselves than they would have when programming everything on their own, four disagreed, and two were undecided.

Asked about the greatest benefit of the AI-SDLC, respondents clearly valued the structure and clear procedure most (8 of 12 substantive answers); early error detection in Specify and Design, confidence in code quality, and team collaboration followed far behind (3 mentions each). In checking AI-generated code, respondents focused on behaviour rather than on the code itself: trying the software out in operation was named most often (5 mentions), followed by automated tests and CI, while manual code reading was named no more often than cross-checking with a second agent or model (3 each). This fits the low rating for following the agent's code. Pull request review appears only once in the free text, although seven respondents had agreed on the closed item that they used it, suggesting it played a smaller role in practice than the ratings imply.

The suggestions for improvement fall into a small number of themes. Three respondents asked for a stronger or faster focus on tests, two for an explicit lightweight path for small tasks so that Specify and Design are not skipped ad hoc, and two for an additional step covering security and post-deployment monitoring. Individual respondents proposed specifying UI requirements as testable scenarios during the Design phase, reducing the initial effort of the Bootstrap phase through automation, and changes to team organisation. Two respondents would not change anything. Finally, the recommendation item shows a consistently positive, if not enthusiastic, picture: answers range from 7 to 10 with a median of 8, five respondents fall into the promoter range and none into the detractor range.

\section{Conclusion and Future Work}

The AI-SDLC was received positively by this cohort. Its clarity and the structure of the phases were the most frequently named benefit, and the repository-local instruction files were judged effective: 11 of 12 respondents agreed that they helped the agent adhere to project conventions, and not one respondent agreed that they had little noticeable effect. The rating items and the free-text accounts do not fully agree, however. Among the respondents who described a concrete case, more reported situations in which the instructions did not help than situations in which they did: conventions the agent forgot, an instruction file that grew large enough to consume too much context, and rules that had to be added only after a problem occurred. Repository-local instructions therefore appear to work, but neither automatically nor without maintenance.

The clearest open issue concerns the relationship between confidence and comprehension. Respondents reported confidence in the correctness of the generated code, yet the statement that they could follow and check that code at any time received the lowest agreement of any positively worded item in the survey, and 11 of 13 described a situation in which they had adopted code without fully understanding it. Verification also relied more on trying the software out in operation (5 mentions) than on automated tests and continuous integration (3 mentions). This points to two places where the process should be strengthened: the test structure before deployment, and an explicit step that supports understanding generated code rather than merely validating its functionality.

Three concrete improvements follow from the answers. First, artifacts and repository-local configuration files are not updated automatically. One respondent described in detail how this breaks down in teamwork: a change agreed in one developer's session is known to that developer's agent but not to the others', so decisions about implementation, architecture, and prioritization do not reach the whole team. Keeping specifications current is therefore a collaboration problem, not only a documentation one. This is particularly relevant to AI-SDLC, where agents can operate autonomously within lifecycle phases while developers retain control through reviewing and modifying persistent project artifacts. Second, architectural decisions were rarely recorded in structured form; ADRs were named by only one respondent. Although AI-SDLC provides ADRs for consequential architectural decisions, their role should therefore be made more explicit in practice. Third, one respondent had defined an own ``new-feature'' skill for a recurring task. Together with requests for a more lightweight process, this suggests a shorter path for small changes without requiring the full cycle.

Two limitations qualify these results. The cohort had little prior software engineering experience---five beginners and seven with basic knowledge, none very experienced---so most respondents had no structured process of their own against which to compare the AI-SDLC, and their assessment may be correspondingly favorable. With 13 respondents from a single cohort, the results describe this group and are not generalizable beyond it.

Future work follows directly from these points. The improvements outlined above should be implemented and the survey repeated with a later cohort, so that the changes can be assessed against a comparable measurement and the data basis broadened. A cohort with more professional software engineering experience would be particularly informative, since those participants could compare the AI-SDLC against a process they already know. Beyond that, measuring outcomes independently of self-report---code quality, test coverage, lifecycle adherence, or review activity in the project repositories---would address the most fundamental limitation of the present design, namely that the results reported here concern perceived rather than directly observed behavior.\\

\noindent\textbf{Declaration on Generative AI Use.}
Generative AI assisted with language editing and Python implementation. Study design, analysis, interpretation, and verification remained with the authors.

\bibliographystyle{splncs}
\bibliography{references, manualreferences}

@article{weijters2012misresponse,
    author  = {Weijters, Bert and Baumgartner, Hans},
    title   = {Misresponse to Reversed and Negated Items in Surveys: A Review},
    journal = {Journal of Marketing Research},
    volume  = {49},
    number  = {5},
    pages   = {737--747},                                                                                                                                         
    year    = {2012},
    doi     = {10.1509/jmr.11.0368}
  }

@article{carroll2007fidelity,                                                                                                                                   
    author  = {Carroll, Christopher and Patterson, Malcolm and Wood, Stephen and
               Booth, Andrew and Rick, Jo and Balain, Shashi},
    title   = {A Conceptual Framework for Implementation Fidelity},
    journal = {Implementation Science},
    volume  = {2},
    number  = {1},                                                                                                                                                
    pages   = {40},
    year    = {2007},
    doi     = {10.1186/1748-5908-2-40}
  }

@article{davis1989perceived,                                                                                                                                    
    author  = {Davis, Fred D.},
    title   = {Perceived Usefulness, Perceived Ease of Use, and User Acceptance of
               Information Technology},
    journal = {MIS Quarterly},
    volume  = {13},
    number  = {3},
    pages   = {319--340},                                                                                                                                         
    year    = {1989},
    doi     = {10.2307/249008}
  }

@article{reichheld2003one,
    author  = {Reichheld, Frederick F.},
    title   = {The One Number You Need to Grow},
    journal = {Harvard Business Review},
    volume  = {81},
    number  = {12},                                                                                                                                               
    pages   = {46--54},
    year    = {2003}
  }

@book{knaster2020safe,
  author    = {Knaster, Richard and Leffingwell, Dean},
  title     = {{SAFe} 5.0 Distilled: Achieving Business Agility with the Scaled Agile Framework},
  publisher = {Addison-Wesley},
  address   = {Boston, MA},
  year      = {2020},
  isbn      = {978-0-13-682039-0}
}

@article{piskala_spec-driven_2026,
	title = {Spec-{Driven} {Development}: {From} {Code} to {Contract} in the {Age} of {AI} {Coding} {Assistants}},
	volume = {abs/2602.00180},
	url = {https://consensus.app/papers/specdriven-developmentfrom-code-to-contract-in-the-age-of-piskala/9f08ebb29bcf5a85b1126e6b427cd249/},
	doi = {10.48550/arxiv.2602.00180},
	journal = {ArXiv},
	author = {Piskala, Deepak Babu},
	month = jan,
	year = {2026},
}

@article{watanabe_use_2026,
	title = {On the {Use} of {Agentic} {Coding}: {An} {Empirical} {Study} of {Pull} {Requests} on {GitHub}},
	url = {https://consensus.app/papers/on-the-use-of-agentic-coding-an-empirical-study-of-pull-watanabe-li/4e69ad3f723e50b98ad34aa0131a2a41/},
	doi = {10.1145/3798166},
	journal = {ACM Transactions on Software Engineering and Methodology},
	author = {Watanabe, Miku and Li, Hao and Kashiwa, Yutaro and Reid, Brittany and Iida, Hajimu and Hassan, Ahmed},
	month = oct,
	year = {2026},
}

@misc{martin_ai-sdlc_2026,
	title = {The {AI}-{SDLC} {Template} and {Method} {Pre}-{Release}},
	copyright = {Creative Commons Attribution 4.0 International},
	url = {https://zenodo.org/doi/10.5281/zenodo.22724343},
	doi = {10.5281/ZENODO.22724343},
	urldate = {2026-09-12},
	publisher = {Zenodo},
	author = {Martin, Andreas and Schwander, Sandro},
	month = jun,
	year = {2026},
	note = {Version 1.0.0-beta, Zenodo, DOI: 10.5281/ZENODO.22724343, URL: https://zenodo.org/doi/10.5281/zenodo.22724343},
}

@misc{martin_ai-sdlc_2026-1,
	title = {The {AI}-{SDLC} {Template} and {Method}: {AI}-{Assisted} {Software} {Development} {Life} {Cycle}},
	copyright = {Creative Commons Attribution 4.0 International},
	shorttitle = {The {AI}-{SDLC} {Template} and {Method}},
	url = {https://zenodo.org/doi/10.5281/zenodo.22833065},
	doi = {10.5281/ZENODO.22833065},
	urldate = {2026-09-18},
	publisher = {Zenodo},
	author = {Martin, Andreas and Schwander, Sandro},
	month = sep,
	year = {2026},
	note = {Version 1.0.0, Zenodo, DOI: 10.5281/ZENODO.22833065, URL: https://zenodo.org/doi/10.5281/zenodo.22833065},
}

@inproceedings{madani_towards_2025,
	title = {Towards the {Integration} of {Large} {Language} {Models} into the {Software} {Development} {Life} {Cycle}: {A} {Systematic} {Literature} {Review}},
	url = {https://consensus.app/papers/towards-the-integration-of-large-language-models-into-the-madani-neumann/598a77b700475e468fa0ea91e881bdd3/},
	doi = {10.1109/fllm67465.2025.11390990},
	booktitle = {2025 3rd {International} {Conference} on {Foundation} and {Large} {Language} {Models} ({FLLM})},
	author = {Madani, M. and Neumann, Ksenia and Nahhas, A. and Chernigovskaya, Maria and Walia, Damanpreet Singh and Turowski, Klaus},
	month = nov,
	year = {2025},
	pages = {773--781},
}

@misc{agent_skills_agent_2026,
	title = {Agent {Skills}: {A} {Standardized} {Way} to {Give} {AI} {Agents} {New} {Capabilities} and {Expertise}},
	url = {https://agentskills.io/},
	author = {{Agent Skills}},
	year = {2026},
}

@misc{nygard_documenting_2011,
	title = {Documenting {Architecture} {Decisions}},
	url = {https://cognitect.com/blog/2011/11/15/documenting-architecture-decisions},
	author = {Nygard, Michael},
	month = nov,
	year = {2011},
}

@misc{agentsmd_agentsmd_2026,
	title = {{AGENTS}.md: {A} {Simple}, {Open} {Format} for {Guiding} {Coding} {Agents}},
	url = {https://agents.md/},
	author = {{AGENTS.md}},
	year = {2026},
}

@misc{github_specification-driven_2025,
	title = {Specification-{Driven} {Development} ({SDD})},
	url = {https://github.com/github/spec-kit/blob/main/spec-driven.md},
	author = {{GitHub}},
	year = {2025},
	note = {Published: GitHub repository github/spec-kit},
}

@misc{bockeler_understanding_2025,
	title = {Understanding {Spec}-{Driven}-{Development}: {Kiro}, spec-kit, and {Tessl}},
	url = {https://martinfowler.com/articles/exploring-gen-ai/sdd-3-tools.html},
	author = {Böckeler, Birgitta},
	month = oct,
	year = {2025},
	note = {Series: Exploring Gen AI
Published: martinfowler.com},
}

@book{leffingwell_scaling_2007,
	address = {Boston, MA},
	series = {Agile {Software} {Development} {Series}},
	title = {Scaling {Software} {Agility}: {Best} {Practices} for {Large} {Enterprises}},
	isbn = {978-0-321-45819-3},
	publisher = {Addison-Wesley},
	author = {Leffingwell, Dean},
	year = {2007},
}

@book{anderson_kanban_2010,
	address = {Sequim, WA},
	title = {Kanban: {Successful} {Evolutionary} {Change} for {Your} {Technology} {Business}},
	isbn = {978-0-9845214-0-1},
	publisher = {Blue Hole Press},
	author = {Anderson, David J.},
	year = {2010},
}

@book{forsgren_accelerate_2018,
	address = {Portland, OR},
	title = {Accelerate: {The} {Science} of {Lean} {Software} and {DevOps}: {Building} and {Scaling} {High} {Performing} {Technology} {Organizations}},
	isbn = {978-1-942788-35-5},
	publisher = {IT Revolution Press},
	author = {Forsgren, Nicole and Humble, Jez and Kim, Gene},
	year = {2018},
}

@book{humble_continuous_2010,
	address = {Boston, MA},
	series = {Addison-{Wesley} {Signature} {Series} ({Fowler})},
	title = {Continuous {Delivery}: {Reliable} {Software} {Releases} through {Build}, {Test}, and {Deployment} {Automation}},
	isbn = {978-0-321-60191-9},
	publisher = {Addison-Wesley},
	author = {Humble, Jez and Farley, David},
	year = {2010},
}

@misc{fowler_continuous_2006,
	title = {Continuous {Integration}},
	url = {https://martinfowler.com/articles/continuousIntegration.html},
	author = {Fowler, Martin and Foemmel, Matthew},
	year = {2006},
	note = {Published: martinfowler.com},
}

@book{martin_clean_2017,
	address = {Boston, MA},
	title = {Clean {Architecture}: {A} {Craftsman}'s {Guide} to {Software} {Structure} and {Design}},
	isbn = {978-0-13-449416-6},
	publisher = {Prentice Hall},
	author = {Martin, Robert C.},
	year = {2017},
}

@misc{fowler_test_2012,
	title = {Test {Pyramid}},
	url = {https://martinfowler.com/bliki/TestPyramid.html},
	author = {Fowler, Martin},
	month = may,
	year = {2012},
	note = {Published: martinfowler.com},
}

@book{beck_test-driven_2003,
	address = {Boston, MA},
	title = {Test-{Driven} {Development}: {By} {Example}},
	isbn = {978-0-321-14653-3},
	publisher = {Addison-Wesley},
	author = {Beck, Kent},
	year = {2003},
}

@misc{gloaguen_evaluating_2026,
	title = {Evaluating {AGENTS}.md: {Are} {Repository}-{Level} {Context} {Files} {Helpful} for {Coding} {Agents}?},
	shorttitle = {Evaluating {AGENTS}.md},
	url = {http://arxiv.org/abs/2602.11988},
	doi = {10.48550/arXiv.2602.11988},
	urldate = {2026-09-14},
	publisher = {arXiv},
	author = {Gloaguen, Thibaud and Mündler, Niels and Müller, Mark and Raychev, Veselin and Vechev, Martin},
	month = jun,
	year = {2026},
	note = {arXiv:2602.11988 [cs.SE]},
}

@article{abrahao_software_2025,
	title = {Software {Engineering} by and for {Humans} in an {AI} {Era}},
	volume = {34},
	url = {https://consensus.app/papers/software-engineering-by-and-for-humans-in-an-ai-era-abraho-grundy/dcd930f0ccf752bd9ba3f77e21d1e762/},
	doi = {10.1145/3715111},
	journal = {ACM Transactions on Software Engineering and Methodology},
	author = {Abrahão, S. and Grundy, John and Pezzè, Mauro and Storey, M. and Tamburri, D.},
	month = feb,
	year = {2025},
	pages = {1--46},
}

@article{dong_survey_2025,
	title = {A {Survey} on {Code} {Generation} with {LLM}-based {Agents}},
	volume = {abs/2508.00083},
	url = {https://consensus.app/papers/a-survey-on-code-generation-with-llmbased-agents-dong-jiang/b29c7815e88a5dd6b23ba097efc2b20f/},
	doi = {10.48550/arxiv.2508.00083},
	journal = {ArXiv},
	author = {Dong, Yihong and Jiang, Xue and Qian, Jiaru and Wang, Tian and Zhang, Kechi and Jin, Zhi and Li, Ge},
	month = jul,
	year = {2025},
}

@article{zhao_is_2025,
	title = {Is {Vibe} {Coding} {Safe}? {Benchmarking} {Vulnerability} of {Agent}-{Generated} {Code} in {Real}-{World} {Tasks}},
	volume = {abs/2512.03262},
	url = {https://consensus.app/papers/is-vibe-coding-safe-benchmarking-vulnerability-of-zhao-wang/bf570b6ce6405afa8f89f8abb9981fa4/},
	doi = {10.48550/arxiv.2512.03262},
	journal = {ArXiv},
	author = {Zhao, Songwen and Wang, Danqing and Zhang, Kexun and Luo, Jiaxuan and Li, Zhuo and Li, Lei},
	month = dec,
	year = {2025},
}

@article{wang_software_2025,
	title = {Software {Development} {Life} {Cycle} {Perspective}: {A} {Survey} of {Benchmarks} for {Code} {Large} {Language} {Models} and {Agents}},
	volume = {abs/2505.05283},
	url = {https://consensus.app/papers/software-development-life-cycle-perspective-a-survey-of-wang-li/a61a457d25645987880692eba94e82d2/},
	doi = {10.48550/arxiv.2505.05283},
	journal = {ArXiv},
	author = {Wang, Kaixin and Li, Tianlin and Zhang, Xiaoyu and Wang, Chong and Sun, Weisong and Liu, Yang and Shi, Bin},
	month = may,
	year = {2025},
}

@article{sapkota_vibe_2025,
	title = {Vibe {Coding} vs. {Agentic} {Coding}: {Fundamentals} and {Practical} {Implications} of {Agentic} {AI}},
	volume = {abs/2505.19443},
	url = {https://consensus.app/papers/vibe-coding-vs-agentic-coding-fundamentals-and-practical-sapkota-roumeliotis/cd98f66602e45dd49ee0777a83840c21/},
	doi = {10.48550/arxiv.2505.19443},
	journal = {ArXiv},
	author = {Sapkota, Ranjan and Roumeliotis, Konstantinos and Karkee, Manoj},
	month = may,
	year = {2025},
}
\end{document}